\documentclass[showpacs,preprintnumbers,amsmath,amssymb,twocolumn,superscriptaddress,prb]{revtex4}
\usepackage{graphicx}
\usepackage{amsfonts}
\usepackage{amsmath, amsthm, amssymb}
\usepackage{dsfont}
\usepackage{times}
\usepackage{subfigure}

\newcommand{\e}[1]{\mathrm{e}^{#1}}
\newcommand{\bq}{\begin{equation}}
\newcommand{\eq}{\end{equation}}

\newcommand{\g}{\underline{\gamma}}
\newcommand{\gt}{\underline{\tilde{\gamma}}}
\newcommand{\N}{\underline{\mathcal{N}}}
\newcommand{\Nt}{\underline{\tilde{\mathcal{N}}}}

\newcommand{\D}{\underline{\mathcal{D}}}
\newcommand{\Dt}{\underline{\tilde{\mathcal{D}}}}

\newcommand{\vecr}{\boldsymbol{r}}

\newcommand{\eg}{\textit{e.g. }}%[syn: f.eks., for example, for instance]

\def\i{\mathrm{i}}

\begin{document}
\title[Spin-polarized Josephson current in S$\mid$F$\mid$S junctions with inhomogeneous magnetization]{Spin-polarized Josephson current in S$\mid$F$\mid$S junctions with inhomogeneous magnetization}

\author{Mohammad Alidoust}
\affiliation{Department of Physics, Faculty of Sciences, University of Isfahan, Hezar Jerib Avenue,
Isfahan 81746-73441, Iran}

\author{Jacob Linder}
\affiliation{Department of Physics, Norwegian University of
Science and Technology, N-7491 Trondheim, Norway}

\author{Gholamreza Rashedi}
\affiliation{Department of Physics, Faculty of Sciences, University of Isfahan, Hezar Jerib Avenue,
Isfahan 81746-73441, Iran}

\author{Takehito Yokoyama}
\affiliation{Department of Applied Physics, University of Tokyo, Tokyo 113-8656, Japan}

\author{Asle Sudb{\o}}
\affiliation{Department of Physics, Norwegian University of
Science and Technology, N-7491 Trondheim, Norway}

\date{Received \today}

\begin{abstract}
We study numerically the properties of spin- and charge-transport
in a current-biased nanoscale diffusive
superconductor$\mid$ferromagnet$\mid$superconductor junction when
the magnetization texture is non-uniform. Specifically, we
incorporate the presence of a Bloch/Neel domain walls and conical
ferromagnetism, including the role of spin-active interfaces. The
superconducting leads are assumed to be of the conventional
$s$-wave type. In particular, we investigate how the 0-$\pi$
transition is influenced by the inhomogeneous magnetization
texture and focus on the particular case where the charge-current
vanishes while the spin-current is non-zero. In the case of a
Bloch/Neel domain-wall, the spin-current can be seen only for one
component of the spin polarization, whereas in the case of conical
ferromagnetism the spin-current has the three components. This is in
contrast to a scenario with a homogeneous exchange field, where
the spin-current vanishes completely. We explain all of these
results in terms of the interplay between the triplet anomalous
Green's function induced in the ferromagnetic region and the local
direction of the magnetization vector in the ferromagnet.
Interestingly, we find that the spin-current exhibits
discontinuous jumps at the 0--$\pi$ transition points of the
critical charge-current. This is seen both in the presence of a
domain wall and for conical ferromagnetism. We explain this result
in terms of the different symmetry obeyed by the current-phase
relation when comparing the charge- and spin-current.
Specifically, we find that whereas the charge-current obeys the
well-known relation $I_c(\phi) = -I_c(2\pi-\phi)$, the
spin-current satisfies $I_s(\phi) = I_s(2\pi-\phi)$, where $\phi$
is the superconducting phase difference.
\end{abstract}

 \maketitle

\section{Introduction}
Because of the interesting phenomena that
superconductor$\mid$ferromagnet$\mid$superconductor
(S$\mid$F$\mid$S) structures exhibit, including their potential
applications in spintronics\cite{wolf,prinz} and quantum
computing\cite{spintro,quantuminfo,dimainfo,Golubov}, this field
of research is presently studied
extensively.\cite{Golubov,BuzdinRev} Usual electronic devices are
based on the properties of flowing electrons through circuits,
whereas spintronic devices are based on direction and number of
flowing spins. In many spintronics devices, like magnetic
tunnelling junctions, spin polarized currents are generated when
an imbalance between spin up and down carriers occurs. This
imbalance can arise \textit{e.g.} by using magnetic materials or
applying a magnetic field. The discovery of the giant
magnetoresistance (GMR) effect \cite{GMR} today forms the
basis of the leading technology for information storage by
magnetic disc drives. Spin coupling and its advantageous high
speeds at very low powers\cite{kikkawa} of these devices promise
applications for logic and storage
applications.\cite{mucciolo,governale,wang}

The possibility of a $\pi$-state in a S$\mid$F$\mid$S systems was
predicted theoretically in Refs. \cite{Bulaevskii,Buzdin1} and has
been observed experimentally.\cite{ryazanov} Near such a
transition point, the junction ground state energy has two minima
versus $\phi$ at $\phi=0$ and $\phi=\pi$. The coexistence of
stable and metastable 0 and $\pi$ states in the transition zone
can produce two flux peaks for one external quantum flux in
superconducting quantum interference device (SQUID)-like geometry,
and renders the system a qubit.\cite{zr} The characteristic length
of the ferromagnetic layer where the first $0$--$\pi $ transition
occurs is of the order of the magnetic coherence length $\xi
_{F}$. In the dirty limit, that is achievable in most of the
experimentally studied S$\mid$F$\mid$S structures. Here, $\xi _{F}$ is
given by $\sqrt{D/h}$ where $D$ denotes the diffusion constant and
$h$ is the magnitude of ferromagnetic exchange field. Therefore,
the experimental observation of such 0-$\pi$ transitions in
nanoscale devices requires a low exchange energy $h$. Such
conditions were achieved using weak ferromagnetic CuNi or PdNi
alloys. The experimental observations of the critical
charge-current oscillations shows such 0-$\pi$ transitions as a
function of the ferromagnet thickness and
temperature.\cite{kontos,ryazanov,ryazanov2,obonzov} The
consequence of the exchange splitting at the Fermi
level\cite{demler}is that the Cooper pairs wave function shows
damped oscillations in the ferromagnet, resulting in the
appearance of the well known $\pi$-state in S$\mid$F$\mid$S
systems.\cite{Bulaevskii} In contrast to the usual 0-state in
superconductor-normal metal-superconductor junctions, the phase
shift equal to $\pi$ across the junction in the ground state
reverses the direction of the supercurrent,\cite{ryazanov} and
considerably changes the density of states (DOS) in the F
metal.\cite{kontos} The $\pi$-states can also be observed in
nonmagnetic junctions of high-$T_c$
superconductors\cite{vanharlingen} and in non-equilibrium
nanoscale superconducting structures.\cite{baselmans} 

In the
ballistic limit, the transport properties of a S$\mid$F$\mid$S
junction can be understood on a microscopic level in terms of
Andreev bound-states.\cite{andreev} The 0--$\pi$ transition
is then due to the spin dependence of the Andreev bound 
states.\cite{Tanaka 97} Because of the averaging of the
quasiclassical Green's function \cite{serene} over momentum
directions, the relevant equations simplify in the dirty transport
regime. This averaging of Green's function can be understood by
noting that in the presence of impurities and scattering centers,
the direction of motion of electrons are random and physical
quantities should be averaged over all directions. This averaging
is valid as long as the mean free path of the diffusive layer is
much smaller than length scales of the system that are
superconducting coherence length $\xi_{S}=\sqrt{D/\Delta_0}$ and
the decay length of Cooper pairs wave functions inside ferromagnet
$\xi_{F}=\sqrt{D/h}$, where $\Delta_0$ is the superconducting
order parameter. The charge-current $I_c(\phi)$ and the local DOS
are two principle quantities that are strongly influenced by the
proximity effect. These two quantities were studied for various
geometries by using of quasiclassical Green's function in clean
and dirty limits in several works, \textit{e.g.}
\cite{golubov,bergeret,zareyan,mohammadkhani_prb_06,
linder_prb_07,yokoyama_prb_07, valls, eschrig, fominov,
fogelstrom, cuoco_prb_09}.

Up to now, the majority of works studying S$\mid$F$\mid$S
junctions have considered a homogeneous exchange field in the
ferromagnet, including half-metallic ferromagnets.\cite{Eschrig2,Asano,Takahashi,Eschrig3,Grein} In the presence of inhomogeneous magnetization
textures, several new effects have been predicted in the
literature including the possibility of a long-range triplet
component. Such an inhomogeneous magnetization texture may be
created artificially by setting up several layers of ferromagnets
with misaligned magnetizations.\cite{barash_prb_02,
pajovic_prb_06, houzet_prb_07, crouzy_prb_07, sperstad_prb_08,
popovic_arxiv_09} Alternatively, inhomogeneous magnetization may
arise naturally in the presence of domain walls or non-trivial
patterns for the local ferromagnetic moment. An example of the
latter is the conical ferromagnet Ho. Very recently, two
theoretical studies have predicted qualitatively new effects in
S$\mid$F and S$\mid$F$\mid$S hybrid structures where F
is a conical ferromagnet \cite{linder_prb_09, halasz_prb_09}. Due
to the inhomogeneous nature of the magnetization in Ho, the
spin-properties of the proximity-induced superconducting
correlations are expected to undergo a qualitative change compared
to the case of homogeneous ferromagnetism. Such changes may also
be expected in the domain-wall case. A more realistic modeling of
hybrid structures involving superconductors and ferromagnets
demands that such non-trivial magnetization textures and also the
spin-dependent properties of the interface regions \cite{Hernando,
Cottet_1} are taken into account seriously. It was recently shown that the
latter may induce qualitatively new features in the local DOS of
S$\mid$F layers \cite{cottet_prb_09} and S$\mid$N layers with
magnetic interfaces \cite{linder_prl_09}.

Another consequence of inhomogeneous magnetization, be it in the form of multiple misaligned layers or intrinsic non-uniformity within a single ferromagnetic layer, is that the Josephson current should become spin-polarized. This has been noted by several authors in the context of superconductors coexisting with helimagnetic or spiral magnetic order \cite{kulic_prb_01, eremin_prb_06} as well as ferromagnetic superconductors \cite{linder_prb_07b, zhao_prb_06}.
However, the spin-polarization of the Josephson current has not been studied in the arguably simplest case of a single ferromagnetic layer with inhomogeneous magnetization contacted by two conventional $s$-wave superconductors.

To this end, we develop in this paper a model for an S$\mid$F$\mid$S junction where both inhomogeneous magnetization and spin-active interfaces are incorporated, and then proceed to solve the problem numerically. More specifically, we will investigate variations of spin- and
charge- currents versus changing of the thickness of F layer
$d_{F}$ for a hybrid S$\mid$F$\mid$S structure with $s$-wave superconductors. \textit{We find that a spin-current flows through the junction whenever the magnetization is inhomogeneous, and that it features discontinuous jumps whenever the junction undergoes a 0--$\pi$ transition.}
We compare these variations for three types of magnetization
textures \textit{i.e.}, homogeneous, domain wall, and a conical exchange field. \textit{We also show that for certain values of $d_{F}$, the critical charge-current vanishes whereas a pure spin-current flows through the system.} Moreover, we demonstrate how it is possible to obtain a pure spin-current by tuning the phase difference between the superconductors.

\section{Theory}\label{sec:theory}

To investigate the behavior of the ferromagnetic Josephson
junction, we employ a full numerical solution of the
quasiclassical equations of superconductivity \cite{serene} in the
diffusive limit \cite{usadel}, which allows us to access the full
proximity effect \cite{mcmillan_pr_68} regime. Importantly,
we also take into account the spin-dependent phase-shifts (spin-DIPS)
microscopically\cite{Cottet_2} that are present at the superconductor$\mid$ferromagnet interfaces.
For the purpose
of stable and efficient numerical calculations, it is convenient to
employ the Ricatti-parametrization of the Green's function as follows:
\cite{hammer_prb_07,schopohl_prb_95,konstandin_prb_05}
\begin{align}\label{eq:g}
\hat{g} &= \begin{pmatrix}
\N(\underline{1}-\g\gt) & 2\N\g \\
2\Nt\gt & \Nt(-\underline{1} + \gt\g) \\
\end{pmatrix}.
\end{align}
Here, $\hat{g}^2=\hat{1}$ since
\begin{align}
\N=(1+\g\gt)^{-1}\; \Nt = (1+\gt\g)^{-1}.
\end{align}
We use $\underline{\ldots}$ for $2\times2$ matrices and $\hat{\ldots}$ for
$4\times4$ matrices.
In order to calculate the Green's function $\hat{g}$, we need to solve the
Usadel equation \cite{usadel} with appropriate boundary conditions at $x=-d_F/2$
and $x=d_F/2$.
We introduce the superconducting coherence length as $\xi_S = \sqrt{D_S/\Delta_0}$.
Following the notation of Ref. \cite{linder_prb_09}, the Usadel equation reads
\begin{align}\label{eq:usadel}
D\partial(\hat{g}\partial\hat{g}) + \i[\varepsilon\hat{\rho}_3 + \text{diag}
[\boldsymbol{h}\cdot\underline{\boldsymbol{\sigma}},(\boldsymbol{h}\cdot
\underline{\boldsymbol{\sigma}})^\mathcal{T}], \hat{g}]=0,
\end{align}
and we employ the following realistic boundary conditions for all our computations in this paper: \cite{Hernando}
\begin{align}\label{eq:bc}
2\zeta d_F\hat{g} \partial \hat{g} = [\hat{g}_\text{BCS}(\phi), \hat{g}] +
\i (G_S/G_T) [\text{diag}(\underline{\tau_3}, \underline{\tau_3}), \hat{g}]
\end{align}
at $x=-d_F/2$. Here, $\partial \equiv \frac{\partial}{\partial x}$ and we defined
$\zeta=R_B/R_F$
as the ratio between the resistance of the barrier region and the resistance
in the ferromagnetic film. The barrier conductance is given by $G_T$,
whereas the parameter $G_S$ describes the spin-DIPS taking place at the F
side of the interface where the magnetization is assumed to parallel to the $z$-axis. The boundary condition at $x=d_F/2$ is obtained by letting
$G_S \to (-\tilde{G}_S)$ and $\hat{g}_\text{BCS}(\phi) \to [-\hat{g}_\text{BCS}
(-\phi)]$ in Eq. (\ref{eq:bc}), where
\begin{align}
\g_\text{BCS}(\phi) &= \i\underline{\tau_2}s/
(1+c)\e{\i\phi/2},\notag\\
\gt_\text{BCS}(\phi) &= \g_\text{BCS}(\phi)\e{-\i\phi}.
\end{align}
Above, $\tilde{G}_S$ is allowed to be different from $G_S$ in general. For instance,
if the exchange field has opposite direction at the two interfaces due to the presence of a domain wall,
one finds $\tilde{G}_S=-G_S$. The total superconducting phase difference is $\phi$, and we have defined
$s=\sinh(\vartheta),
c=\cosh(\vartheta)$ with $\vartheta=\text{atanh}(\Delta_0/\varepsilon)$
using $\Delta_0$ as the superconducting gap. Note that we use the bulk
solution in the superconducting region, which is a good approximation when
assuming that the superconducting region is much less disordered than
the ferromagnet and when the interface transparency is small, as considered here.
We use units such that $\hbar=k_B=1$.

The values of $G_S$ and $G_T$ may be calculated explicitly from a
microscopic model, which allows one to characterize the
transmission $\{t_{n,\sigma}^j\}$ and reflection amplitudes
$\{r_{n,\sigma}^j\}$ on the $j\in\{S,F\}$ side. Under the
assumption of tunnel contacts and a weak ferromagnet, one obtains
with a Dirac-like barrier model\cite{Cottet_1,Cottet_2, Hernando}
\begin{align}
G_T = G_Q\sum_n T_n,\; G_S = 2G_Q\sum_n\Big( \rho_n^F - \frac{4\tau_n^S}{T_n}\Big)
\end{align}
upon defining $T_n = \sum_\sigma |t_{n,\sigma}^S|^2$ and
\begin{align}
\rho_n^F = \text{Im}\{r_{n,\uparrow}^F (r_{n,\downarrow}^F)^*\},\; \tau_n^S = \text{Im}\{t_{n,\uparrow}^S (t_{n,\downarrow}^S)^*\}.
\end{align}
For simplicity, we assume that the interface is characterized by $N$ identical scattering channels. Omitting the subscript '$n$',
the scattering coefficients are obtained as:
\begin{align}
r_\sigma^F &= (k_\sigma^F - k_\sigma^S - \i k_\sigma^SZ_\sigma)/\mathcal{D}_\sigma,\notag\\
t_\sigma^S &= 2\sqrt{k_\sigma^S k_\sigma^F}/\mathcal{D}_\sigma,
\end{align}
with the definitions $\mathcal{D}_\sigma = k_\sigma^S + k_\sigma^F + \i k_\sigma^SZ_\sigma$ and
\begin{align}
k_\sigma^S = \sqrt{2m_S\mu_S},\; k_\sigma^F = \sqrt{2m_F(\mu_F + \sigma h)}.
\end{align}
Here, $Z_\sigma = Z_0 + \sigma Z_S$ is the spin-dependent barrier potential. Defining the polarization $P=h/\mu_F$ in the ferromagnet
and the polarization $\nu = Z_S/Z_0$ for the barrier, we will set $P=\nu$.

\begin{figure}
\includegraphics[width=7cm]{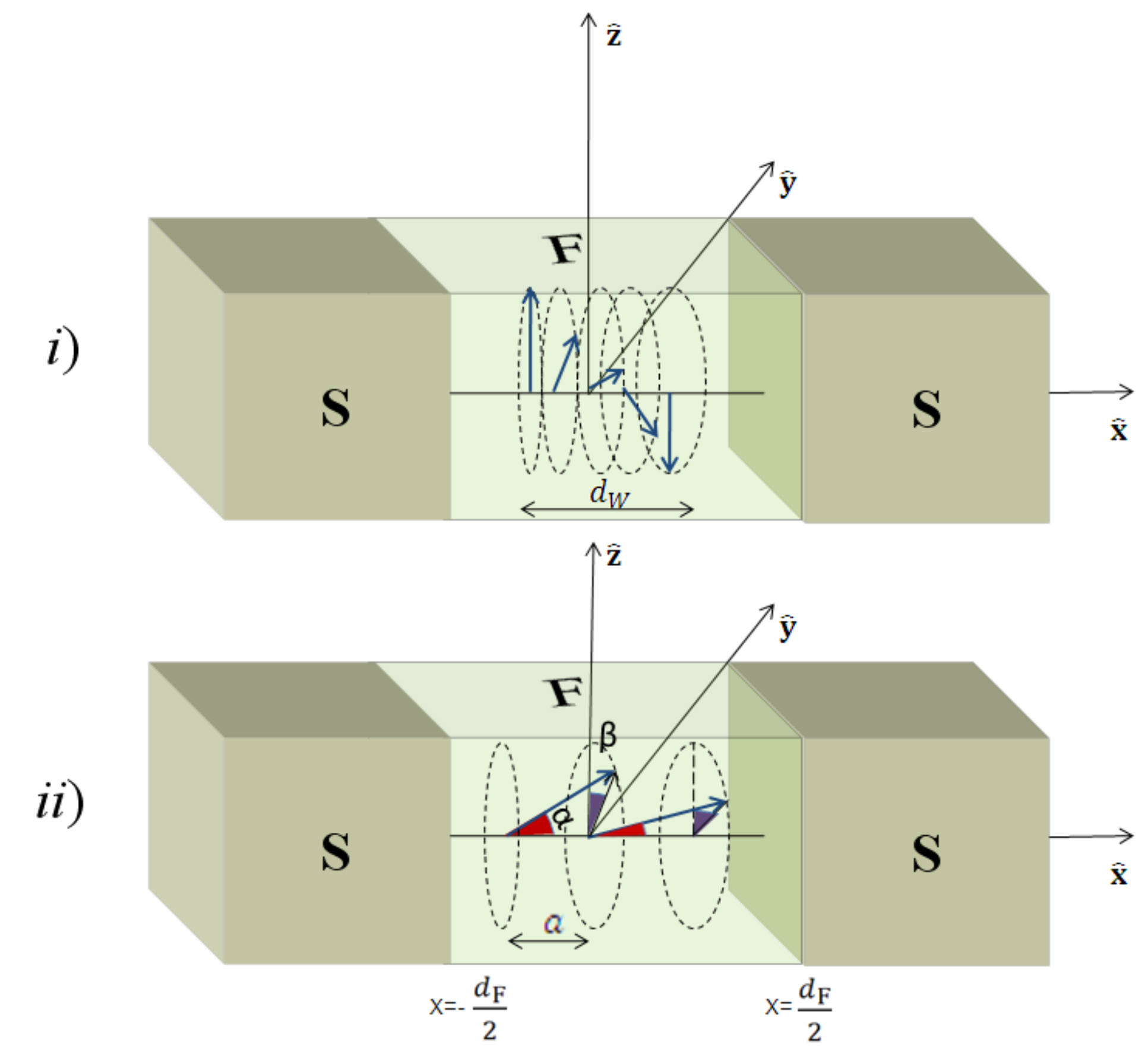}
\caption{\label{fig:model} (Color online) \textit{i)} The S$\mid$F$\mid$S
junction with Bloch domain wall of ferromagnet and \textit{ii)}
with conical type of ferromagnet. The magnetization texture for the Neel wall is obtained by replacing the $x$-component of the magnetization with an $y$-component in case \textit{i)}. The blue
arrows show the magnetic moments in F layer. The magnetic moment
for Bloch/Neel domain wall has two components and for conical type has
three components.}
\end{figure}

In this paper we will consider three types of inhomogeneous
magnetic textures: Bloch, N\'{e}el and a conical structure. These
structures are all different from a homogenous magnetic texture.
The first two types of magnetic textures are assumed to be
located at the center of the F layer. The Bloch model is
demonstrated by $\mathbf{h}=h(\cos\theta \hat{y}+\sin\theta
\hat{z})$ and its structure is shown in Fig. \ref{fig:model}.
Similarly, the N\'{e}el model reads $\mathbf{h}=h(\cos\theta
\hat{x}+\sin\theta\hat{z})$ where we defined $\theta$ as
follows\cite{konstandin_prb_05}:
\begin{align}
\theta=-\arctan(x/d_{W}).
\end{align}
Here, $d_{W}$ is the width of domain wall and we assumed that the
center of F layer is located at the origin, \textit{i.e}, $x=0$ as
shown in Fig. \ref{fig:model}. 

For the conical case, we adopt a model where the magnetic moment rotates on the
surface of a cone with defined apex angle $\alpha$ and turning
angle $\beta$. This structure is shown in Fig.1 ($\alpha$ and
$\beta$ will determine the kind of material in use). If we assume
that the distances of interatomic layers are $a$,\cite{sosnin} the
spiral variation in the exchange field can be written as
\begin{align}
\mathbf{h}=h(\cos\alpha\hat{x}+\sin\alpha[\sin(\beta
x/a)\hat{y}+\cos(\beta x/a)\hat{z}]).
\end{align}

To characterize the transport properties of the system, we define
the normalized charge and spin current densities according to:
\begin{align}
\frac{I_c}{I_{c,0}}&= \Big|\int_0^\infty \text{d}\tilde{\varepsilon}\;\text{Tr}\Big\{\hat{\rho}_3 \Big(\check{g}\frac{\partial\check{g}}{\partial\tilde{x}}\Big)^K\Big\}  \Big|,
\end{align}
and
\begin{align}
\frac{I_s^z}{I_{s,0}}&=\int_0^\infty\text{d}\tilde{\varepsilon}\;\text{Tr}\Big\{\hat{\rho}_3\hat{\tau}_3\Big(\check{g}\frac{\partial\check{g}}{\partial\tilde{x}}\Big)^K\Big\},
\end{align}
respectively, where $\tilde{\varepsilon}=\varepsilon/\Delta_0$, $\tilde{x}=x/d_F$,
and $\hat{\rho}_3=\text{diag}(1,1,-1,-1)$. Here $I_c$ and $I_s^z$
are the charge- and the $z$-component of the spin-current flowing
in the $\hat{x}$-direction, respectively. The normalization constants are:
\begin{align}
I_{c,0} &= \frac{N_0eD\Delta_0}{8d_F},\; I_{s,0} = \frac{\hbar I_{c,0}}{2e},
\end{align}
where $N_0$ is the normal state DOS per spin and $D$ is the diffusion constant.
In general, the
spin-current for other components of spin polarization $j\in
\{x,y,z\}$ is given as:
\begin{align}\label{eq:spincurrent}
\frac{I_s^j}{I_{s,0}}=\int_0^\infty\text{d}\tilde{\varepsilon}\;\text{Tr}\Big\{\hat{\rho}_3\hat{\nu}_j\Big(\check{g}\frac{\partial\check{g}}{\partial\tilde{x}}\Big)^K\Big\}
,\; \hat{\nu}_j = \begin{pmatrix}
\underline{\tau_j} & \underline{0} \\
\underline{0} &\underline{\tau_j}^* \\
\end{pmatrix}.
\end{align}
\par

Above, $\hat{\rho_i}$, $\hat{\tau_i}$, and $\underline{\tau_i}$
are Pauli matrices that are defined in the appendix C and the
reader may consult Appendix B for the derivation of the expression for $I_s^j/I_{s,0}$. Under the assumption of an equilibrium
situation, the Keldysh block of Green's function reads:
\begin{align}
\hat{g}^K = [\hat{g}^R - \hat{g}^A]\tanh(\beta\varepsilon/2),
\end{align}
where $\hat{g}^R$and
$\hat{g}^A=-(\hat{\rho_3}\hat{g}^R\hat{\rho_3})$ are Retarded and
Advanced blocks of $\check{g}$ respectively, and $\beta=1/T$ is
inverse temperature.

\section{Results and Discussion}
We now present our main results of this paper, namely a study of
how the critical currents depend on the thickness $d_F$ of the
junction in the presence of homogeneous and inhomogeneous exchange
field and spin-active interfaces. In order to focus on a realistic
experimental setup, we choose the junction parameters as follows.
For a weak, diffusive ferromagnetic alloy such as Pd$_x$Ni$_{1-x}$,
the exchange field $h/\Delta_0$ is tunable by means of the doping
level $x$ to take values in the range meV to tens of meV. Here,
we will fix $h/\Delta_0=15$, which typically places the exchange
field $h$ in the range 15-25 meV. The thickness $d_F$ of the
junction is allowed to vary in the range $d_F/\xi_S\in[0.5,1.2]$,
which is equivalent to $9-21.6$ nm for a superconducting coherence
length of $\xi_S=18$ nm as can be obtained for \eg Nb. This range
of layer thicknesses $d_F$ are experimentally
feasible.\cite{oboznov_prl_06} The ratio of $G_S/G_T$ is
calculated according to the microscopic expressions given in the
previous section only for uniform and domain wall exchange fields because we will set $G_S=0$ for the conical ferromagnet.
We choose $\mu_F = 1$ eV and $\mu_S = 10$ eV for the Fermi level
in the ferromagnet and superconductor, respectively, and consider a
relatively low barrier transparency of $Z_0=3$. The electron mass
$m_F$ and $m_S$ in both the F and S regions is taken to be the
bare one ($\simeq 0.5$ MeV). Any change in effective mass
translates into an effective barrier resistance due to the
Fermi-wavevector mismatch, which thus is captured by the parameter
$Z_\sigma$. The interface region is assumed to exhibit a much
higher electrical resistance than in the bulk of the ferromagnet,
and we set $\zeta=R_B/R_F =4$. For more stability in our
computations we used the Ricatti parametrization and also inserted
a small imaginary part $\delta=5\times10^{-3}\Delta_0$ in the
quasiparticle energy $\varepsilon$, effectively modeling inelastic
scattering. A considerable amount of CPU-time was put into the
calculations of the current, as we solved for a fine mesh of both
quasiparticle energies $\varepsilon$ and phase differences $\phi$
for each value of the width $d_F$. As will be discussed in detail
below, we find that for S$\mid$F$\mid$S structures with
spin-singlet $s$-wave superconducting leads, a spin-current exists
only for domain wall structures and conical type of the
ferromagnet layer, whereas it vanishes completely in the case of a
homogeneous exchange field. Both the charge- and spin-current are
evaluated in the middle of the F region, $x=0$. The charge-current
is conserved throughout the system, and its magnitude is thus
independent of $x$. The spin-current, on the other hand, is not
conserved and in fact suffers a depletion close to the S$\mid$F
interfaces and vanishes completely in the superconducting regions.
The critical charge-current is given by
$I_{cc}=$max$_\phi\{I_c(\phi)\}$, and the phase giving the
critical current may be denoted $\phi_c$. We define the critical
spin-current as $I_{cs}$=$I_s(\phi_c)$, which means that we are
effectively considering \textit{the spin-polarization of the
critical charge-current}, which should be the most sensible choice
physically in a current-biased scenario. Note that this is different from the maximum value of
the spin-current as a function of $\phi$. 

\subsection{Critical currents vs. thickness $d_F$ for homogeneous exchange field}
\begin{figure}
\includegraphics[width=9.0cm]{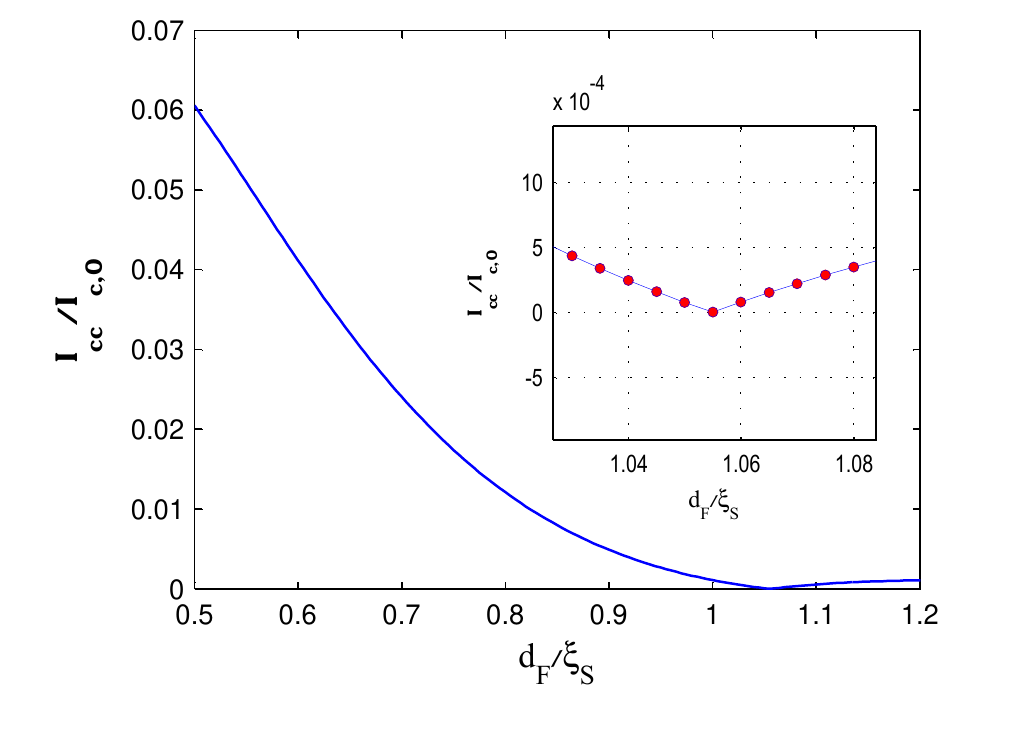}
\caption{\label{fig:homogeneous} (Color online) The variation of the
normalized critical charge-current versus the
thickness of a homogeneous F layer. The inset panel zooms in on the behavior near the 0-$\pi$ transition. As long as the exchange field is constant, we find that the spin-current $I_s$ vanishes.}
\end{figure}

First, we consider how the charge- and spin-currents are influenced by
changing the thickness of F layer $d_F$ in the homogeneous
magnetic texture case. We fix the temperature at $T/T_c=0.2$, and use the
microscopic expression for spin-DIPS $G_\phi$ at the two
boundaries. The result is shown in the Fig. \ref{fig:homogeneous}. The critical charge-current in the region of
$d_{F}$ from 0.5$\xi_{S}$ to 1.2$\xi_{S}$ vanishes at one point.
This point is the first 0-$\pi$ transition point. We
found that, for all strengths of the exchange field and
spin-DIPS, the spin-current $I_s$ is zero. Unlike
the case of spin-triplet superconductors, we can not see any spin-current even
for $\hat{x}$ and $\hat{y}$ directions of spin
polarization\cite{Rashedi,Asano2}. In fact, one can confirm this finding analytically for all
components of spin polarization at least for linearized Usadel equation 
and transparent boundaries. The reason for the vanishing spin-current will become clear from the discussion in the following section, when noting that only the $S_z=0$ odd-frequency triplet and even-frequency singlet components are induced by the proximity effect in the ferromagnetic region.

\subsection{Critical currents vs. thickness $d_F$ for Bloch and N\'{e}el domain walls }

We now turn our attention to the first example of a non-trivial magnetization texture in the ferromagnet, namely the scenario of a Bloch or Neel domain wall. We use the same values for $h$ and $T$ as in the previous section, and set the domain wall width $d_W$ to $d_W/d_F = 0.5$ and assume that it is centered in the ferromagnet. Although the domain-wall structure dictates that the magnetization is not fully directed along the $z$-axis at the interfaces, we have verified numerically that the influence of the spin-DIPS parameter $G_S$ is negligible for our choice of parameters, such that we still can use the boundary conditions in Sec. \ref{sec:theory}.

\begin{figure}
\includegraphics[width=9cm]{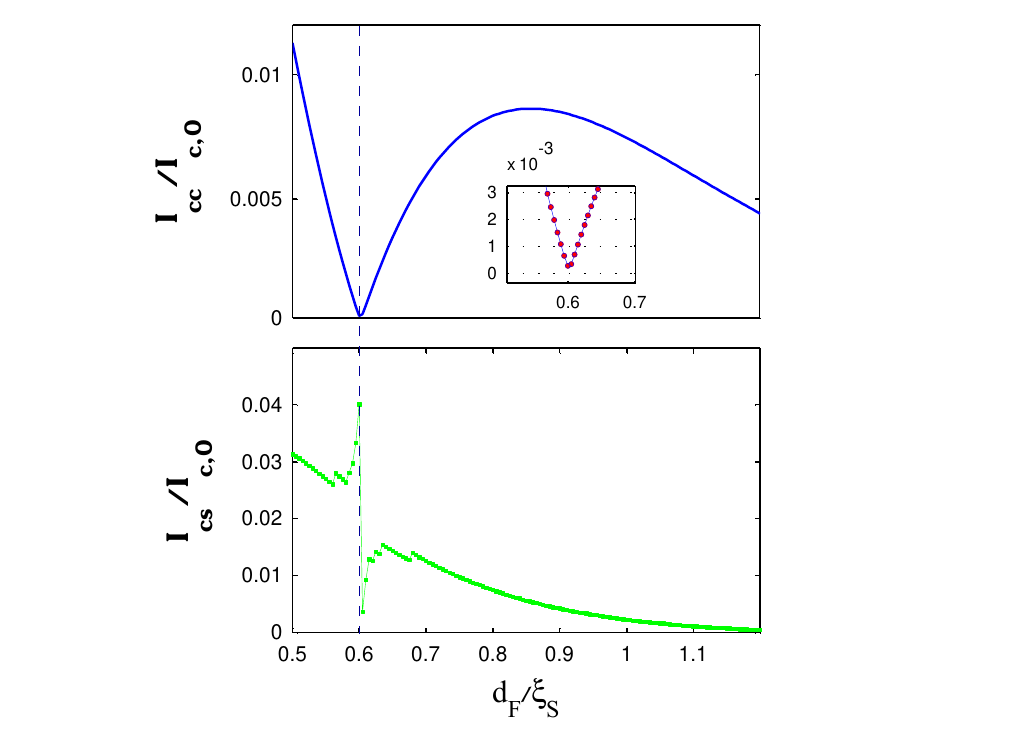}
\caption{\label{fig:domainwall_width} (Color online) The variations of
normalized critical spin- and charge-currents vs. increasing the thickness $d_F$ of F layer with a Bloch domain wall structure.}
\end{figure}

The results of the variation of the normalized critical spin- and
charge-currents vs. $d_{F}/\xi_{S}$ are shown in the Fig.
\ref{fig:domainwall_width} , considering here a Bloch wall
texture. Contrary to the homogeneous case considered in the previous section, we now see that \textit{a finite spin-current flows through the system}. For this type of magnetization
texture, we note that the spin-current only exists for one
component of the spin polarization: the $\hat{x}$-component. Only one component of the spin-current would be present also in the Neel domain wall case, as we shall explain below. The spin-current features a discontinuous jump at the
same value of the thickness where the charge-current undergoes a
0--$\pi$ transition, namely $d_F/\xi_S \simeq 0.6$. For this
value of thickness the spin-current has a rapid variation. We have
checked numerically with a very high resolution of $d_F$ (a step
of $5\times10^{-4}$ for $d_F/\xi_S$) that this result does not
pertain to noise or any error. For this type of magnetization
texture, we also note that a spin-current only exists for one
component of the spin polarization: the $\hat{x}$-component.  We
will explain the reason for both the presence of such jumps in the
spin-current and the polarization properties later.

We now explain why only one component of the spin-polarization is present both in the Bloch and Neel domain wall case. In order to understand the reason for this, it  is instructive to consider the interplay between the triplet anomalous Green's function $\boldsymbol{f}$, given by
\begin{align}
\boldsymbol{f} = \Big[\frac{f_\downarrow-f_\uparrow}{2}, -\frac{i(f_\downarrow+f_\uparrow)}{2}, \frac{f_{\uparrow\downarrow} + f_{\downarrow\uparrow}}{2}\Big]
\end{align}
and the local direction of the exchange field $\boldsymbol{h}$. In S$\mid$F proximity structures, $\boldsymbol{f}$ tends to align as much as possible with $\boldsymbol{h}$. For a homogeneous exchange field $\boldsymbol{h}$ in the $z$-direction, one thus obtains that only the $S_z=0$ opposite-spin pairing triplet component $f_t = \boldsymbol{f}_z$ is present, as is well-known. Consider now the Bloch domain wall case. The $\boldsymbol{f}$-vector then contains only $y$- and $z$-components. Now, the spin expectation value of the Cooper pair is provided by
\begin{align}
\langle \boldsymbol{S} \rangle \propto \i ( \boldsymbol{f} \times \boldsymbol{f}^*),
\end{align}
and we immediately infer that only a spin-polarization in the
$x$-direction will be present. A similar line of reasoning for the
Neel domain wall case leads to the result that only a
spin-polarization in the $y$-direction is present. Since we are
evaluating the spin-current in the middle of the F region, the
$z$-component of the local exchange field is absent there. In that
case, $\langle \boldsymbol{S} \rangle$ should equal to zero
according to our argument above. The reason for why a finite
spin-current is nevertheless obtained must be attributed to a lag
between the $\boldsymbol{f}$ and $\boldsymbol{h}$ vectors, such
that they do not follow each other exactly. One would expect that
for a slower variation of the local exchange field, the lag would
decrease.

\subsection{Critical currents vs. thickness $d_F$ for conical type of magnetization texture}

\begin{figure}
\includegraphics[width=0.6\textwidth]{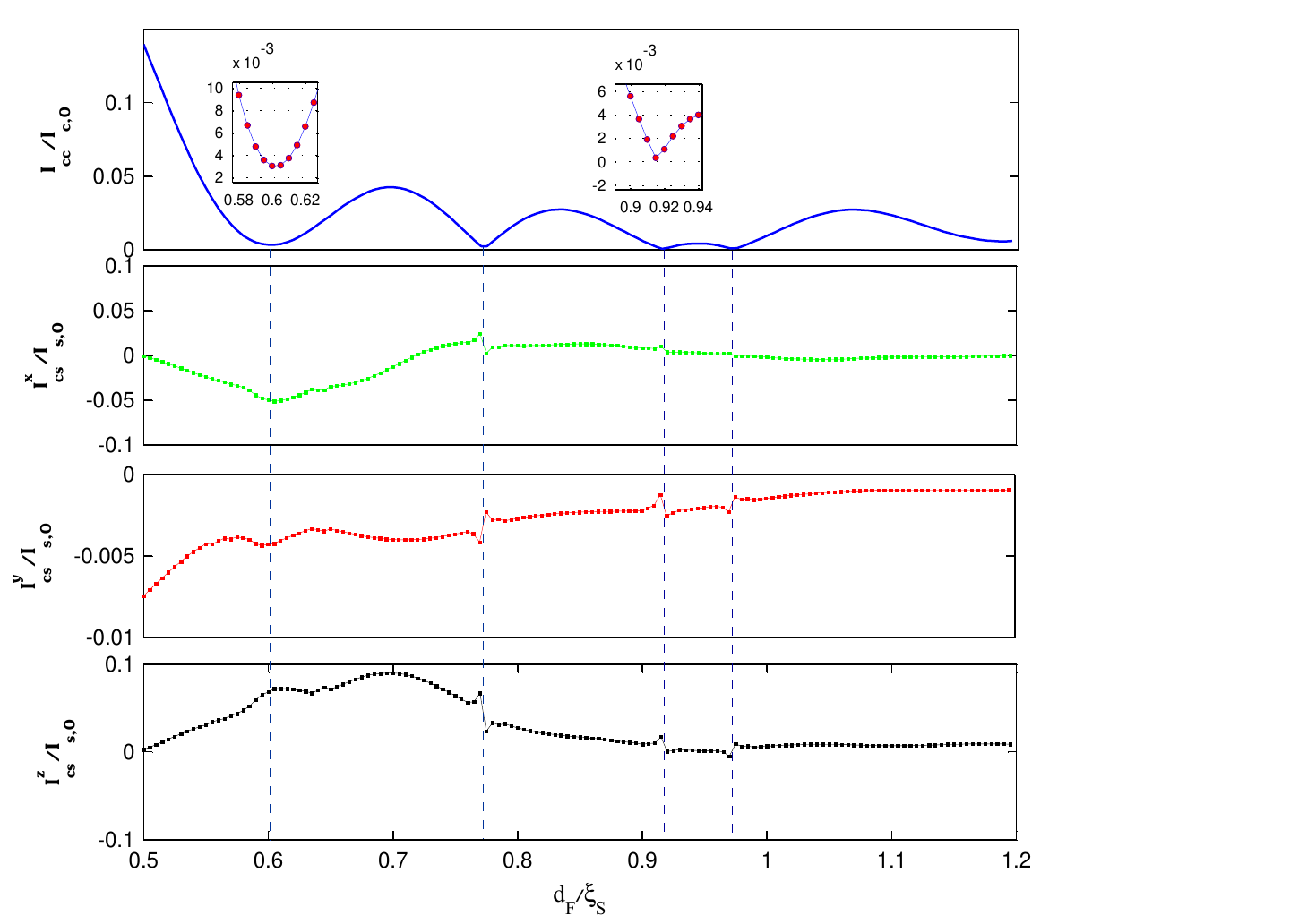}
\caption{\label{fig:conical_width} (Color online) Plot of the normalized critical spin- and charge-current vs. the normalized
thickness $d_F$ of the F layer for a conical type of magnetization texture. All three components of the spin
polarization have a considerable magnitude in the entire range of $d_F$ considered here. }
\end{figure}

Finally, we turn our attention to the conical model for magnetization, relevant to Ho. For simplicity, we set the
$G_S=0$ at the two boundaries at $-d_F/2$ and $d_/2$.  The distance between the atomic layers $a$ is equal to
0.02$d_{F}$, $\alpha=4\pi/9$, and rotating angle $\beta=\pi/6$ per
interatomic layer. These values of $a$, $\alpha$ and $\beta$ are chosen based on the actual lattice parameters of Ho. The result of the investigation of how the critical spin- and charge-currents vary as a function of $d_{F}/\xi_{S}$
is shown in the Fig. \ref{fig:conical_width}. In this case, we see a qualitatively new behavior for the charge-current as compared to the previous two subsections where we treated a homogeneous exchange field and a domain-wall ferromagnet, respectively. In Fig. \ref{fig:conical_width}, one observes a superimposed pattern of fast oscillations on top of the usual 0--$\pi$ oscillations, which are slower. This is in agreement with the very recent work by Halasz \textit{et al.} \cite{halasz_prb_09}, who also reported the generation of rapid oscillations on top of the conventional 0--$\pi$ transitions of the current in the weak-proximity effect regime. These faster oscillations pertain to the inhomogeneous magnetization texture considered here, although they are not seen in the domain-wall case. This fact indicates that they are sensitive to the precise form of the magnetization structure in the ferromagnet, and that they do not appear simply as a result of a general inhomogeneity.

As can be seen in the Fig. \ref{fig:conical_width}, the critical-charge current
has five local minima, out of which three are 0--$\pi$
transition points. In Fig. \ref{fig:conical_width}, the first dotted horizontal line indictates a minima which is irrelevant to a 0--$\pi$ transition, whereas the three following dotted lines indicate minima which correspond to such transitions. The last local minima is located near $d_F/\xi_S=1.2$ and is not indicated by a dotted line in Fig. \ref{fig:conical_width}. This is in contrast to the homogeneous and domain wall case, where only one 0--$\pi$ transition point is seen in the range of $d_F$ considered here. As for the spin-current, the behavior is similar to the Bloch
wall structure, with a rapid
variation at the transition point. As mentioned previously, we
have investigated these discontinuous jumps of the spin-current with a very high resolution for $d_F$ to ensure that do not stem from numerical errors or noise. We now proceed to an explanation for this effect.

\subsection{Origin of the discontinuous jumps in the spin-current}\label{sec:explanation}

\begin{figure}
\includegraphics[width=0.5\textwidth]{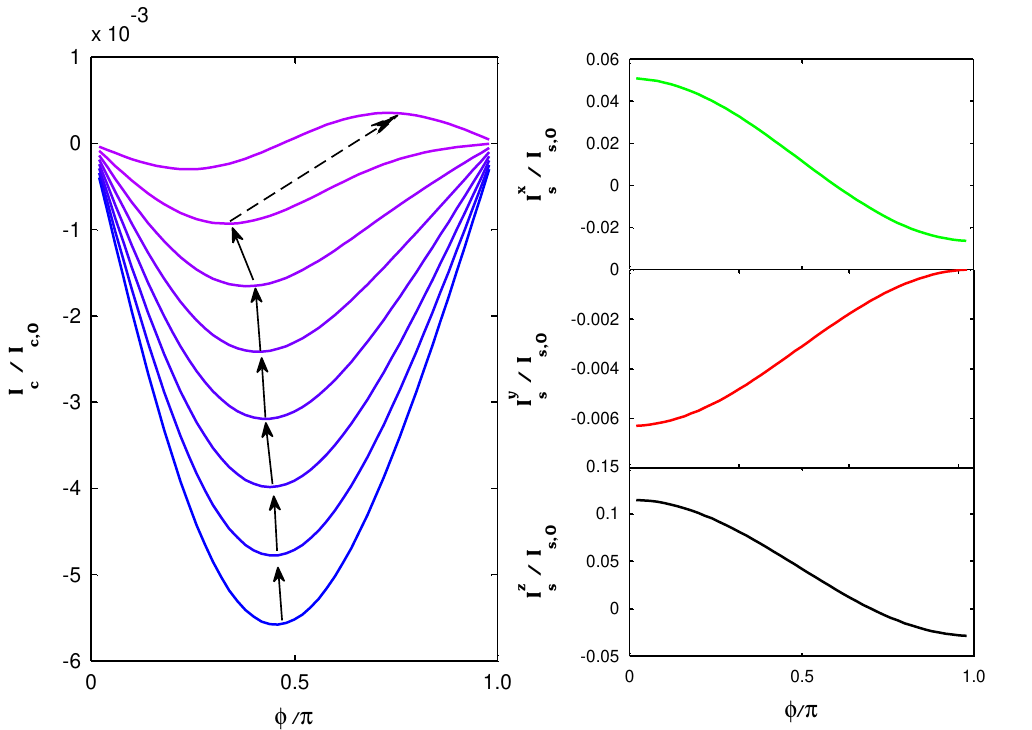}
\caption{\label{fig:conical_phase} (Color online) Plot of the current-phase relation for the charge- and spin-currents. We are considering a conical magnetization texture, and the curves range from $d_F/\xi_S=0.7655$ to $0.7725$ in steps of $1\times10^{-3}$ in the charge-current panel. For the spin-current panels, the variation of the current-phase relation upon changing $d_F$ is negligible and we give results only for $d_F/\xi_S=0.7725$. Note that the \textit{critical} spin-current nevertheless varies strongly with $d_F$ near the transition points as shown in Fig. \ref{fig:conical_width}, since the critical phase displays a strong dependence on $d_F$ in this regime.}
\end{figure}

In order to understand the mechanism behind the discontinuous jumps of the spin-current near the 0--$\pi$ transition of the junction, we revert briefly to the original definition of the critical spin-current. It is defined as $I_s(\phi_c)$ where $\phi_c$ is the value of the superconducting phase difference which gives the maximum (absolute) value of the charge-current. In effect, the critical spin-current is the spin-polarization of the critical charge-current, which is \textit{distinct} from the maximum value of the spin-current. We now consider in detail the current-phase relation for both charge- and spin-transport near the transition point located at $d_F/\xi_S\simeq 0.772$ (see Fig. \ref{fig:conical_width}). The result for the current-phase relation is shown in Fig. \ref{fig:conical_phase}, where we consider several values of $d_F$ near the transition point. From bottom to top, they range from $d_F/\xi_S=0.7655$ to $0.7725$ in steps of $1\times10^{-3}$. A key point is that we have verified numerically that the charge-current is antisymmetric with respect to $\phi=\pi$ whereas the spin-current is symmetric around this value. More specifically, whereas
\begin{align}
I_c(\phi) = -I_c(2\pi-\phi)
\end{align}
we find numerically that the spin-current satisfies
\begin{align}
I_s(\phi) = I_s(2\pi-\phi)
\end{align}
This is consistent with the finding of Ref. \cite{Rashedi} where transport between spin-triplet superconductors has been investigated. As a result, it suffices to restrict our attention to the range $\phi\in[0,\pi]$. Next, we note that the charge-current is nearly sinusoidal to begin with (bottom curves of Fig. \ref{fig:conical_phase}). Upon increasing $d_F$, and thus approaching the transition point, higher harmonics in the current-phase relation become more protrudent for the charge-current. However, the spin-current remains virtually unafffected by an increase in $d_F$, and we plot the result only for $d_F/\xi_S=0.7725$. Upon increasing $d_F$, the critical phase $\phi_c$ moves away from $\pi/2$ to lower values due to the presence of higher harmonics in the current-phase relation. At the transition point, the phase jumps in a discontinuous manner to $\phi_c>\pi/2$ (dotted arrow in Fig. \ref{fig:conical_phase}). Now, the charge-current has a similar magnitude (in absolute-value) for this new value of $\phi_c$. The spin-current, on the other hand, has a different symmetry with respect to $\phi$ as seen in Fig. \ref{fig:conical_phase}, and varies less rapidly with $d_F$. 
%For instance, it does not vanish at $\phi=\{0,\pi\}$. 
Therefore, the spin-polarization of the current makes a discontinuous jump at the transition point.

\section{Summary}

In summary, we have considered the transport of charge and spin in a nanoscale S$\mid$F$\mid$S Josephson junction when the magnetization texture is inhomogeneous in the ferromagnetic layer.
More specifically, we have investigated how charge and spin Josephson currents are affected by the presence of Bloch/Neel domain walls and conical ferromagnetism, including also the spin-active properties of the interfaces. \textit{We find that a spin-current flows through the junction whenever the magnetization is inhomogeneous, and that it features discontinuous jumps whenever the junction undergoes a 0--$\pi$ transition.} In the case of a Bloch/Neel domain-wall, the spin-current can be seen
only for one component of the spin polarization (the component perpendicular to  both the local direction of the exchange field and that of its derivative), whereas in the case of conical ferromagnetism the spin-current has three
components. For a homogeneous exchange field, the spin-current vanishes. We explain the polarization properties of the spin-current by considering interplay between the triplet anomalous Green's functions induced in the ferromagnetic region and the local direction of the magnetization vector in the ferromagnet. Moreover, we show how the discontinuous jumps in the spin-current stem from the different symmetries for the current-phase relation when comparing the charge- and spin-current. While the charge-current obeys the well-known relation $I_c(\phi) = -I_c(2\pi-\phi)$, the spin-current satisfies $I_s(\phi) = I_s(2\pi-\phi)$, where $\phi$ is the superconducting phase difference. Our results open up new perspectives for applications in superspintronics by exploiting Josephson junctions with non-homogeneous ferromagnets.

\acknowledgments
J.L. thanks I. B. Sperstad for helpful discussions.
J.L. and A.S. were supported by the Research Council of Norway, Grants No. 158518/432 and
No. 158547/431 (NANOMAT), and Grant No. 167498/V30 (STORFORSK).
T.Y. acknowledges support by JSPS.

\appendix
\section{Useful relations for the Green's function}
\noindent By introducing the auxiliary quantities
\begin{align}
\partial_x \N = -\N \D \N,\; \partial_x\Nt = -\Nt \Dt \Nt,
\end{align}
where we have defined
\begin{align}
\D = (\partial_x \g)\gt + \g(\partial_x \gt),\; \Dt = (\partial_x\gt)\g + \gt(\partial_x\g),
\end{align}
we find that the matrix derivative of the Green's function $\partial_x \hat{g}^R$ has the following components:
\begin{align}
(\partial_x \hat{g}^R)_{11} &= -\D\N - (\underline{1}-\g\gt)\N\D\N,\notag\\
(\partial_x \hat{g}^R)_{12} &=2(\partial_x\g)\Nt - 2\g\Nt\Dt\Nt,\notag\\
(\partial_x \hat{g}^R)_{21} &=2(\partial_x\gt)\N - 2\gt\N\D\N,\notag\\
(\partial_x \hat{g}^R)_{22} &=\Dt\Nt + (\underline{1}-\gt\g)\Nt\Dt\Nt
\end{align}
The indices above refer to particle-hole space, and each of the above elements is thus a $2\times2$ matrix in spin-space.

\section{Quasiclassical equation for the spin-current}
\noindent We here show how the matrix structure in the analytical expression Eq. (\ref{eq:spincurrent}) for the spin-current is obtained in the quasiclassical approximation. The starting point is the quantum mechanical expression for the expectation value of the spin-current:
\begin{align}\label{eq:spincurrent_basic}
\langle \boldsymbol{j}_S(\vecr) \rangle = \frac{1}{2m} \langle \text{Im}\{ \Psi^\dag(\vecr) \nabla_{\vecr} \text{diag}(\boldsymbol{\sigma}, \boldsymbol{\sigma}^*)\Psi(\vecr)\}\rangle,
\end{align}
with a fermion operator basis $\Psi$ given as
\begin{align}
\Psi^\dag(\vecr) = (\psi_\uparrow^\dag(\vecr), \psi_\downarrow^\dag(\vecr), \psi_\uparrow(\vecr), \psi_\downarrow(\vecr)).
\end{align}
Above, $\boldsymbol{\sigma}$ is the Pauli matrix vector. It should be noted that the spin-current $\boldsymbol{j}_S$ is a tensor since it has a flow-direction in real space in addition to a polarization in spin-space. For clarity, we consider in what follows the $\sigma_2$-component corresponding to the polarization in the $\hat{\boldsymbol{y}}$-direction, as an example. We then get from Eq. (\ref{eq:spincurrent_basic}) [using that $\text{Im}\{\i z\} = \text{Re}\{z\}$ for a complex number $z$]
\begin{align}\label{eq:tempapp}
\langle \boldsymbol{j}_S^y(\vecr) \rangle &= \frac{1}{2m}\text{Re}\{-\langle \psi_\uparrow^\dag(\vecr)\nabla_{\vecr}\psi_\downarrow(\vecr)\rangle + \langle \psi_\downarrow^\dag(\vecr)\nabla_{\vecr}\psi_\uparrow(\vecr) \rangle \notag\\
&+ \langle \psi_\uparrow(\vecr)\nabla_{\vecr}\psi_\downarrow^\dag(\vecr) \rangle - \langle \psi_\downarrow(\vecr)\nabla_{\vecr}\psi_\uparrow^\dag(\vecr) \rangle\}\notag\\
&= \frac{1}{4m} \lim_{\vecr\to\vecr'} (\nabla_{\vecr} - \nabla_{\vecr'})[\langle \psi_\uparrow^\dag(\vecr)\psi_\downarrow(\vecr')\rangle \notag\\
&- \langle \psi_\downarrow^\dag(\vecr)\psi_\uparrow(\vecr')\rangle + \langle \psi_\downarrow(\vecr)\psi_\uparrow^\dag(\vecr')\rangle \notag\\
&-\langle \psi_\uparrow(\vecr)\psi_\downarrow^\dag(\vecr')\rangle].
\end{align}
Using the notation of Ref. \cite{morten_diplom}, we define the following representation for the Keldysh Green's function:
\begin{align}
\Big(\hat{G}^K(\vecr,\vecr')\Big)_{mn} &= -\i\sum_j(\hat{\rho}_3)_{mj} \Big\langle [\Psi(\vecr)_j,\Psi^\dag(\vecr')_n]_-\Big\rangle.
\end{align}
It then follows from anticommutation that \eg:
\begin{align}
\langle \psi^\dag_\uparrow(\vecr)\psi_\downarrow(\vecr') \rangle &= \frac{1}{2} \langle \psi^\dag_\uparrow(\vecr)\psi_\downarrow(\vecr') \rangle + \frac{1}{2} \langle \psi^\dag_\uparrow(\vecr)\psi_\downarrow(\vecr') \rangle\notag\\
&= -\frac{1}{2} \langle \psi_\downarrow(\vecr')\psi^\dag_\uparrow(\vecr) \rangle+ \frac{1}{2}\langle \psi^\dag_\uparrow(\vecr)\psi_\downarrow(\vecr') \rangle.
\end{align}
In this way, we can rewrite the last lines of Eq. (\ref{eq:tempapp}) as:
\begin{align}
\langle \boldsymbol{j}_S^y(\vecr) \rangle &= \frac{1}{8m}\lim_{\vecr\to\vecr'} (\nabla_{\vecr} - \nabla_{\vecr'})[\i \Big(\hat{G}^K(\vecr,\vecr')\Big)_{21} \notag\\
&- \i \Big(\hat{G}^K(\vecr,\vecr')\Big)_{12} -\i \Big(\hat{G}^K(\vecr,\vecr')\Big)_{34} \notag\\
&+ \i \Big(\hat{G}^K(\vecr,\vecr')\Big)_{43}]\notag\\
&= -\frac{1}{8m} \lim_{\vecr\to\vecr'} (\nabla_{\vecr} - \nabla_{\vecr'}) \text{Tr}\{ \hat{\rho}_3\notag\\
&\times\text{diag}(\underline{\tau_2},\underline{\tau_2}^*)\times \hat{G}^K(\vecr,\vecr')\}.
\end{align}
For the $x$ and $z$-components, one replaces $\underline{\tau_2}$ with $\underline{\tau_1}$ and $\underline{\tau_3}$, respectively.

\section{Pauli Matrices}
The Pauli matrices that are used in this paper are

\begin{align}
\underline{\tau_1} &= \begin{pmatrix}
0 & 1\\
1 & 0\\
\end{pmatrix},\;
\underline{\tau_2} = \begin{pmatrix}
0 & -\i\\
\i & 0\\
\end{pmatrix},\;
\underline{\tau_3} = \begin{pmatrix}
1& 0\\
0& -1\\
\end{pmatrix},\notag\\
\underline{1} &= \begin{pmatrix}
1 & 0\\
0 & 1\\
\end{pmatrix},\;
\hat{1} = \begin{pmatrix}
\underline{1} & \underline{0} \\
\underline{0} & \underline{1} \\
\end{pmatrix},\;
\hat{\tau}_i = \begin{pmatrix}
\underline{\tau_i} & \underline{0}\\
\underline{0} & \underline{\tau_i} \\
\end{pmatrix},\notag\\
\hat{\rho}_1 &= \begin{pmatrix}
\underline{0} & \underline{\tau_1}\\
\underline{\tau_1} & \underline{0} \\
\end{pmatrix},\;
\hat{\rho}_2 =  \begin{pmatrix}
\underline{0} & -\i\underline{\tau_1}\\
\i\underline{\tau_1} & \underline{0} \\
\end{pmatrix},\;
\hat{\rho}_3 = \begin{pmatrix}
\underline{1} & \underline{0}\\
\underline{0} & -\underline{1}  \\
\end{pmatrix}.
\end{align}


\begin{thebibliography}{99}

\bibitem{wolf} S. A. Wolf, D. D. Awschalom, R. A. Buhrman, J. M. Daughton, S. von Molnar, M. L. Roukes, A. Y. Chtchelkanova, D. M. Treger , Science \textbf{294}, 1488 (2001).

\bibitem{prinz} G. A. Prinz, Science \textbf{282}, 1660 (1998).

\bibitem{spintro} I. Zutic, J. Fabian, and S. D. Sarma, Rev. Mod. Phys. \textbf{76}, 323 (2004).

\bibitem{quantuminfo} L. B. Ioffe, V. B. Geshkenbein, M. V. Feigelman, A. L. Fauchere and G. Blatter, Nature (London) \textbf{398}, 679 (1999).

\bibitem{dimainfo} L. B. Ioffe, M. V. Feigel'man, A. Ioselevich, D. Ivanov, M.Troyer, and G. Blatter, Nature (London) \textbf{415}, 503 (2002).

\bibitem{Golubov} A. G. Golubov, M. Yu. Kupriyanov, and E. llichev,
Rev. Mod. Phys. {\bf 76}, 411 (2004).

\bibitem{BuzdinRev} A. I. Buzdin, Rev. Mod. Phys. {\bf 77}, 935 (2005).

\bibitem{GMR} M. N. Baibich, J. M. Broto, A. Fert, F. Nguyen Van Dau,
F. Petroff, P. Etienne, G. Creuzet, A. Friederich, and J.
Chazelas, Phys. Rev. Lett. \textbf{61}, 2472 (1988).

\bibitem{kikkawa} J. M. Kikkawa and D. D. Awschalom, Nature (London) \textbf{397}, 139 (1999).

\bibitem{mucciolo} E. R. Mucciolo, C. Chamon, and C. M. Marcus, Phys. Rev. Lett. \textbf{89}, 146802 (2002).

\bibitem{governale} M. Governale, F. Taddei, and R. Fazio, Phys. Rev. B \textbf{68}, 155324 (2003).

\bibitem{wang} B. Wang, J. Wang, and H. Guo, Phys. Rev. B \textbf{67}, 092408 (2003).

\bibitem{Bulaevskii} L. N. Bulaevskii, V. V. Kuzii, and A. A. Sobyanin, Pis'ma Zh. Eksp. Teor. Fiz. {\bf 25}, 314 (1977) [ JETP Lett. {\bf 25}, 290 (1977)].

\bibitem{Buzdin1} A. I. Buzdin, L. N. Bulaevskii, and S. V. Panyukov, Pis'ma Zh. Eksp. Teor. Fiz. {\bf 35}, 147 (1982) [ JETP Lett. {\bf 35}, 178 (1982)].
\bibitem{ryazanov} V. V. Ryazanov, V. A. Oboznov, A. Yu. Rusanov, A. V. Veretennikov, A. A. Golubov, and J. Aarts, Phys. Rev. Lett. {\bf 86}, 2427 (2001).
\bibitem{zr} Z. Radovi\'c, L. Dobrosavljevi\'c-Gruji\'c, and B. Vuji\v ci\'c, Phys. Rev. B {\bf 63}, 214512 (2001).
\bibitem{kontos} T. Kontos, M. Aprili, J. Lesueur, and X. Grison, Phys. Rev.  Lett. {\bf 86}, 304 (2001); T. Kontos, M. Aprili, J. Lesueur, F. Genet, B. Stephanidis, and R. Boursier, Phys. Rev. Lett. \textbf{89}, 137007 (2002).
\bibitem{ryazanov2} V. V. Ryazanov, V. A. Oboznov, A. S. Prokofiev, V. V. Bolginov, and A. K. Feofanov, J. Low Temp. Phys. \textbf{136}, 385 (2004).

\bibitem{obonzov} V. A. Oboznov, V. V. Bol'ginov, A. K. Feofanov, V. V. Ryazanov,and A. I. Buzdin, Phys. Rev. Lett. \textbf{96}, 197003 (2006).


\bibitem{demler} E. A. Demler, G. B. Arnold, and M. R. Beasley, Phys. Rev. B \textbf{55},15174 (1997).

\bibitem{vanharlingen} D. J. van Harlingen, Rev. Mod. Phys. {\bf 67}, 515 (1995).

\bibitem{baselmans} J. J. A. Baselmans, T. T. Heikkil{\"a}, B. J. van Wees, and T. M. Klapwijk, Phys. Rev. Lett. {\bf 89}, 207002 (2002);
J. J. A. Baselmans, A. F. Morpurgo, B. J. van Wees, and T. M. Klapwijk, Nature (London) {\bf 397}, 43 (1999).

\bibitem{andreev} A. F. Andreev, Zh. Eksp. Teor. Fiz. {\bf 46}, 1823 (1964) [Sov. Phys. JETP {\bf 19}, 1228 (1964)].

\bibitem{Tanaka 97} Y. Tanaka and S. Kashiwaya, Physica {\bf C 274}, 357  (1997).

\bibitem{serene} J. W. Serene and D. Rainer, Phys. Rep. \textbf{101}, 221 (1983).

\bibitem{golubov} A. A. Golubov, M. Yu. Kupriyanov, and Ya. V. Fominov, Pis'ma Zh. \'Eksp. Teor. Fiz. {\bf 75}, 223  (2002) [JETP Lett. {\bf 75}, 190 (2002)].

\bibitem{bergeret} F. S. Bergeret, A. F. Volkov, and K. B. Efetov, Phys. Rev. B {\bf 64}, 134506 (2001); F. Bergeret, A. F. Volkov, and K. B. Efetov, Phys. Rev. B \textbf{68}, 064513 (2003); A. F. Volkov, F. S. Bergeret, and K. B. Efetov, Phys. Rev. Lett. \textbf{90}, 117006 (2003).

\bibitem{zareyan} M. Zareyan, W. Belzig, and Yu. V. Nazarov, Phys. Rev.  Lett. {\bf 86}, 308 (2001).

\bibitem{mohammadkhani_prb_06} G. Mohammadkhani and M. Zareyan, Phys. Rev. B \textbf{73}, 134503 (2006)

\bibitem{linder_prb_07} J. Linder and A. Sudb{\o}, Phys. Rev. B \textbf{75}, 134509 (2007); J. Linder and A. Sudb{\o}, Phys. Rev. B \textbf{76}, 214508 (2007).

\bibitem{yokoyama_prb_07} T. Yokoyama, Y. Tanaka, and A. Golubov, Phys. Rev. B \textbf{72}, 052512 (2005);  Phys. Rev. B \textbf{73}, 094501 (2006);
 Phys. Rev. B \textbf{75}, 134510 (2007);
T. Yokoyama, Y. Tanaka, A. A. Golubov, and Y. Asano, Phys. Rev. B \textbf{73}, 140504(R) (2006); T. Yokoyama, Y. Sawa, Y. Tanaka, and A. A. Golubov, Phys. Rev. B \textbf{75}, 020502(R) (2007);
T. Yokoyama, Y. Tanaka, and A. A. Golubov, Phys. Rev. B \textbf{75}, 094514 (2007); Y. Sawa, T. Yokoyama, Y. Tanaka, and A. A. Golubov, Phys. Rev. B \textbf{75}, 134508 (2007) 
 J. Linder, T. Yokoyama, and A. Sudb{\o}, Phys. Rev. B \textbf{77}, 174507 (2008).

\bibitem{valls}  K. Halterman and O. T. Valls, Phys. Rev. B \textbf{72}, 060514 (2005); K. Halterman, P. H. Barsic, and O. T. Valls, Phys. Rev. Lett. \textbf{99}, 127002 (2007);  P. H. Barsic and O. T. Valls, Phys. Rev. B \textbf{79}, 014502 (2009).

\bibitem{eschrig} T. L{\"o}fwander, T. Champel, and M. Eschrig, Phys. Rev. B 75, 014512 (2007);  J. Kopu, M. Eschrig, J. C. Cuevas, and M. Fogelstr{\"o}m, Phys. Rev. B \textbf{69}, 094501 (2004); T. L{\"o}fwander, T. Champel, J. Durst, and M. Eschrig, Phys. Rev. Lett. \textbf{95}, 187003 (2005); T. Champel, T. L{\"o}fwander, and M. Eschrig, Phys. Rev. Lett. \textbf{100}, 077003 (2008).


\bibitem{fominov}  Ya. V. Fominov, N. M. Chtchelkatchev, and A. A. Golubov, Phys. Rev. B \textbf{66}, 014507 (2002);  Ya. V. Fominov, A. F. Volkov, and K. B. Efetov, Phys. Rev. B \textbf{75}, 104509 (2007)

\bibitem{fogelstrom} M. Fogelstr\"om, Phys. Rev. B {\bf 62}, 11812 (2000).

\bibitem{cuoco_prb_09}  G. Annunziata, M. Cuoco, C. Noce, A. Romano, and P. Gentile, Phys. Rev. B \textbf{80}, 012503 (2009)

\bibitem{Eschrig2} M. Eschrig, J. Kopu, J. C. Cuevas, and G. Sch{\"o}n, Phys. Rev. Lett. \textbf{90}, 137003 (2003).

\bibitem{Asano} Y. Asano, Y. Tanaka, and A. A. Golubov, Phys. Rev. Lett. \textbf{98}, 107002 (2007).

\bibitem{Takahashi} S. Takahashi, S. Hikino, M. Mori, J. Martinek, and S. Maekawa, Phys. Rev. Lett. \textbf{99}, 057003 (2007).

\bibitem{Eschrig3} M. Eschrig and T. L{\"o}fwander, Nat. Phys. \textbf{4}, 138 (2008).

\bibitem{Grein} R. Grein, M. Eschrig, G. Metalidis, and Gerd Sch{\"o}n, Phys. Rev. Lett. \textbf{102} 227005 (2009).

\bibitem{barash_prb_02}  Yu. S. Barash, I. V. Bobkova, and T. Kopp, Phys. Rev. B \textbf{66}, 140503 (2002).

\bibitem{pajovic_prb_06}  Z. Pajovi\'c, M. Bozovi\'c, Z. Radovi\'c, J. Cayssol, and A. Buzdin, Phys. Rev. B \textbf{74}, 184509 (2006)

\bibitem{houzet_prb_07} M. Houzet and A. Buzdin, Phys. Rev. B \textbf{76}, 060504 (2007).

\bibitem{crouzy_prb_07} B. Crouzy, S. Tollis, D. Ivanov, Phys. Rev. B \textbf{75}, 054503 (2007).

\bibitem{sperstad_prb_08} I. B. Sperstad, J. Linder, and A. Sudb{\o}, Phys. Rev. B \textbf{78}, 104509 (2008).

\bibitem{popovic_arxiv_09} Z. Popovi\'c and Z. Radovi\'c, arXiv:0907.2042.

\bibitem{linder_prb_09} J. Linder, T. Yokoyama, and A. Sudb{\o}, Phys. Rev. B \textbf{79}, 054523 (2009).

\bibitem{halasz_prb_09}  G. B. Hal‡sz, J. W. A. Robinson, J. F. Annett, and M. G. Blamire,
Phys. Rev. B \textbf{79}, 224505 (2009).

\bibitem{Hernando} D. Huertas-Hernando, Yu. V. Nazarov, and W. Belzig, Phys. Rev. Lett. \textbf{88}, 047003 (2002); D. Huertas-Hernando, Yu. V. Nazarov, and W. Belzig, arXiv:cond-mat/0204116.

\bibitem{Cottet_1} A. Cottet and W. Belzig, Phys. Rev. B \textbf{72}, 180503 (2005).

\bibitem{cottet_prb_09} A. Cottet and J. Linder, Phys. Rev. B \textbf{79}, 054518 (2009).

\bibitem{linder_prl_09}ÊJ. Linder, T. Yokoyama, A. Sudb{\o}, and M. Eschrig, Phys. Rev. Lett. \textbf{102}, 107008 (2009).



% FMSC spin-current
\bibitem{linder_prb_07b} M. Gr{\o}nsleth, J. Linder, J.-M. B{\o}rven, and A. Sudb{\o}, Phys. Rev. Lett. \textbf{97}, 147002 (2006); J. Linder, M. Gr{\o}nsleth, and A. Sudb{\o}, Phys. Rev. B \textbf{75}, 024508 (2007).

\bibitem{zhao_prb_06} Y. Zhao and R. Shen, Phys. Rev. B \textbf{73}, 214511 (2006).

% helimagnetic superconductor and spiral magnetic order
\bibitem{kulic_prb_01} M. L. Kulic and I. M. Kulic, Phys. Rev. B \textbf{63}, 104503 (2001).

\bibitem{eremin_prb_06} I. Eremin, F. S. Nogueira, and R.-J. Tarento, Phys. Rev. B \textbf{73}, 054507 (2006).





%\bibitem{serene} See \eg J. W. Serene and D. Rainer, Phys. Rep. \textbf{101}, 221 (1983).

\bibitem{usadel} K. Usadel, Phys. Rev. Lett. \textbf{25}, 507 (1970).

\bibitem{mcmillan_pr_68} W. L. McMillan, Phys. Rev. \textbf{175}, 537 (1968).

\bibitem{Cottet_2} A. Cottet, Phys. Rev. B \textbf{76}, 224505 (2007).

\bibitem{hammer_prb_07} J. C. Hammer, J. C. Cuevas, F. S. Bergeret, and W. Belzig, Phys. Rev. B \textbf{76}, 064514 (2007).

\bibitem{schopohl_prb_95} N. Schopohl and K. Maki, Phys. Rev. B \textbf{52}, 490 (1995).

\bibitem{konstandin_prb_05} Alexander Konstandin, Juha Kopu, and Matthias Eschrig, Phys. Rev. B \textbf{72}, 140501 (2005).

\bibitem{sosnin} I. Sosnin, H. Cho, V. T. Petrashov, and A. F. Volkov, Phys. Rev. Lett. \textbf{96}, 157002 (2006).

\bibitem{oboznov_prl_06} V. A. Oboznov, V. V. Bol'ginov, A. K. Feofanov, V. V. Ryazanov, and A. I. Buzdin, Phys. Rev. Lett. \textbf{96}, 197003 (2006).

\bibitem{Rashedi} G. Rashedi, and Yu.A. Kolesnichenko, Physica C \textbf{31-37}, 451 (2007).

\bibitem{Asano2} Y. Asano, Phys. Rev. B \textbf{74}, 220501(R) (2006).


\bibitem{morten_diplom} J. P. Morten, M. Sc. thesis, Norwegian University of Science and Technology (2003).

%\bibitem{wallraff} A. Wallraff, D. I. Schuster, A. Blais, L. Frunzio, J. Majer, M. H. Devoret, S. M. Girvin, and R. J. Schoelkopf, Phys. Rev. Lett. \textbf{95}, 060501 (2005).

%\bibitem{nielsen} M. A. Nielsen and I. L. Chuang, Quantum Computation and Quantum Information (Cambridge University Press, Cambridge England, 2000).

%\bibitem{eilenberger} G. Eilenberger, Z. Phys. 214, 195 (1968).

%\bibitem{daumens} M. Daumens and Y. Ezzahri, Phys. Lett. A \textbf{306}, 344 (2003).

%\bibitem{hirsch} J. E. Hirsch, Phys. Rev. Lett. \textbf{83}, 1834 (1999).

%\bibitem{zhang} S. Zhang, Phys. Rev. Lett. \textbf{85}, 393 (2000).

%\bibitem{bhat} R. D. R. Bhat and J. E. Sipe, Phys. Rev. Lett. \textbf{85}, 5432 (2000).

%\bibitem{hubner} J. Hubner, W. W. Ruhle, M. Klude, D. Hommel, R. D. R. Bhat, J. E. Sipe, and H. M. van Driel, Phys. Rev. Lett. \textbf{90}, 216601 (2003).

%\bibitem{inf-storg} G. A. Prinz, Science \textbf{282}, 1660 (1998).


\end{thebibliography}
\end{document}